\documentclass[sigconf]{acmart}

\usepackage{algorithm}
\usepackage{algorithmic}
\usepackage{float}
\usepackage{amsmath}
\usepackage{amsfonts}
\AtBeginDocument{%
  }

\copyrightyear{2026}
\acmYear{2026}
\setcopyright{cc}
\setcctype{by}
\acmConference[ICCAD '26]{IEEE/ACM International Conference on Computer-Aided Design}{November 08--12, 2026}{San Jose, CA, USA}
\acmBooktitle{IEEE/ACM International Conference on Computer-Aided Design (ICCAD '26), November 08--12, 2026, San Jose, CA, USA}
\acmDOI{10.1145/3831252.3833963}
\acmISBN{979-8-4007-2873-0/2026/11}

\begin{document}

\title{GI-RTL-Fuzz: Grammar-Informed RTL Generation with LLMs for End-to-End Differential Simulator Fuzzing}

\author{Ziyue Zheng}
\affiliation{%
  \institution{Microelectronics Thrust, The Hong Kong University of Science and Technology (Guangzhou)}
  \country{China}
}
\email{zzheng989@connect.hkust-gz.edu.cn}
\author{Zhi Qu}
\authornote{Ziyue Zheng and Zhi Qu contributed equally to this work.}
\affiliation{%
  \institution{Microelectronics Thrust, The Hong Kong University of Science and Technology (Guangzhou)}
  \country{China}
}
\email{zqu901@connect.hkust-gz.edu.cn}
\author{Yangdi Lyu}
\authornote{This work was supported by the National Natural Science Foundation of China under Grant No. 62402412. Corresponding author: yangdilyu@hkust-gz.edu.cn}
\affiliation{%
  \institution{Microelectronics Thrust, The Hong Kong University of Science and Technology (Guangzhou)}
  \country{China}
}
\email{yangdilyu@hkust-gz.edu.cn}

\begin{abstract}
Register-Transfer Level (RTL) simulators are fundamental to modern IC design workflows, yet like any complex software, they frequently contain critical bugs that compromise simulation accuracy. To systematically stress-test these tools, we introduce GI-RTL-Fuzz, a grammar-informed fuzzing framework that leverages Large Language Models (LLMs) to generate high-quality RTL test cases. Our key innovation is guiding the LLM's generation process using Backus-Naur Form (BNF) grammar coverage information, enabling the exploration of complex and rarely-tested language features. By performing differential testing against different RTL simulators, our framework uncovered a significant number of previously unknown inconsistencies in both the linting process and simulation results of major Verilog simulators.
\end{abstract}


\maketitle

\section{Introduction}
As modern System-on-Chip (SoC) designs grow increasingly complex, Hardware Description Languages (HDLs) like SystemVerilog have introduced sophisticated constructs for modeling, synthesis, and verification. Register-Transfer Level (RTL) simulators form the foundation of the Electronic Design Automation (EDA) ecosystem, implicitly trusted by hardware engineers for everything from functional validation to power analysis. However, as the language standards grow increasingly intricate, "silent" simulator bugs ---ranging from subtle semantic discrepancies in event scheduling to catastrophic crashes during compilation---pose a significant risk to the entire hardware lifecycle, potentially leading to multimillion-dollar silicon respins \cite{Manes2021, Saravanan2024}.

HDL standards define syntax using context-free grammars, commonly written in Backus-Naur Form (BNF). For example, a simplified excerpt for Verilog module headers can be written as:
\begin{quote}
\small\ttfamily
module\_decl ::= 'module' id '(' port\_list ')' ';' ... 'endmodule'\\
port\_list ::= $\epsilon$ \;|\; port (',' port)*
\end{quote}
Here, quoted tokens denote terminals, unquoted symbols denote nonterminals, and $\epsilon$ denotes the empty string.

Extensive research in the hardware verification community has focused on discovering vulnerabilities in RTL designs using techniques such as fuzzing~\cite{Laeufer2018, Hur2026, Ragab2026, Zheng2026} and concolic testing~\cite{Tan2025}. However, these methods are primarily designed to verify a Design Under Test (DUT) against a given specification, implicitly assuming the underlying toolchain---including parsers, elaborators, and simulation engines---is bug-free. Recent studies have revealed that even mature, commercially available simulators can exhibit inconsistent behaviors when faced with complex language features or corner-case scheduling scenarios \cite{Zhou2025, Hur2026}. The goal of this paper is to address this critical gap by generating targeted hardware designs to \emph{fuzz} the simulators themselves. Identifying these tool-level bugs is exceptionally challenging due to the massive state space of HDL grammars and the lack of a definitive reference for many advanced SystemVerilog features. 

When testing EDA toolchains---and RTL simulators in particular---traditional fuzzers encounter a significant "input generation bottleneck." Grammar-based fuzzers like Grammarinator \cite{Hodovan2018} and Gramatron \cite{Srivastava2021} excel at generating syntactically correct inputs based on formal grammars (e.g., BNF), but they often lack the semantic awareness required to generate "meaningful" RTL that drives simulators into deep and diverse execution paths. Conversely, mutation-based approaches often struggle to maintain syntactic validity, leading to a high rejection rate by simulator front-ends and limited exploration of deep tool logic \cite{Corradi2024, Vieira2026}.

The emergence of Large Language Models (LLMs) offers a transformative paradigm for RTL generation. LLMs trained on vast repositories of hardware code, such as VeriGen \cite{Thakur2024}, have shown a remarkable ability to synthesize complex Verilog modules from natural language prompts. In the context of simulator fuzzing, these models can act as powerful generators of diverse RTL programs that stress different parts of a simulator's front-end and execution engine. Despite their impressive generative capabilities, LLM-generated code often exhibits syntactic errors, functional inconsistencies, and potential security vulnerabilities \cite{Gadde2026, Fang2025}. Bridging the gap between the unconstrained creativity of LLMs and the rigid requirements of HDL simulation requires a framework that can provide both structural guidance and semantic feedback, with the explicit goal of exposing simulator bugs rather than verifying a fixed hardware design.

To illustrate the severity of these silent bugs, consider a case of illegal operator usage in Verilator. According to the IEEE 1800 standard, reduction NAND (\texttt{\textasciitilde\&}) and reduction NOR (\texttt{\textasciitilde|}) are \emph{unary} operators that can only be applied to a single operand. However, Verilator incorrectly accepts these operators in binary infix positions, treating expressions such as \texttt{var1~\textasciitilde\&~var2} as valid, whereas both Icarus Verilog and ModelSim correctly reject them:
\begin{quote}
\small\ttfamily
// Illegal binary use of reduction operator\\
logic [7:0] var1, var2, result;\\
assign result = var1 ~\textasciitilde\& var2; // Invalid \textasciitilde\&\\

// Expected: Syntax/semantic error\\
// Verilator: [Silent] Incorrect acceptance.\\
// Icarus Verilog/ModelSim: Correct rejection.
\end{quote}
In this scenario, Verilator silently accepts malformed code that violates the language standard, while standards-compliant simulators reject it. Such inconsistencies can lead to designs that appear to work correctly in one simulator but fail silently in others, creating dangerous discrepancies in verification environments.

In this paper, we propose \texttt{GI-RTL-Fuzz}, a grammar-informed RTL generation framework designed specifically for the differential fuzzing of RTL simulators. Our approach leverages the semantic depth of LLMs while enforcing structural constraints through Backus-Naur Form (BNF) grammar guidance. By integrating grammar coverage metrics into the prompting process, \texttt{GI-RTL-Fuzz} systematically directs the LLM to explore rarely-tested "evil" language features that are likely to trigger corner-case bugs in simulator implementations. To improve the syntactic validity of LLM-generated code, we implement an automated self-correcting feedback loop that utilizes simulator error logs to iteratively refine the generated code. Additionally, we establish a robust differential testing pipeline that treats multiple RTL simulators as cross-checking oracles to detect subtle semantic inconsistencies.

The main contributions of this work are as follows:

\begin{itemize}
    \item We develop \texttt{GI-RTL-Fuzz}, a novel RTL \emph{simulator fuzzing} framework that combines the generative power of LLMs with BNF grammar coverage guidance to systematically explore the SystemVerilog feature space from the perspective of simulators.
    \item We introduce an automated self-correcting generation loop that leverages simulator feedback to improve the syntactic validity and semantic richness of LLM-generated RTL test cases, significantly increasing the fraction of inputs that progress deep into simulator execution.
    \item We establish a cycle-accurate differential testing pipeline that treats multiple RTL simulators as fuzzing targets, enabling the discovery of semantic bugs where simulators exhibit inconsistent behaviors.
    \item We demonstrate the effectiveness of our simulator fuzzing framework by identifying 15 previously unreported bugs and inconsistencies in major open-source simulators, including the open-source simulators Verilator~\cite{verilator} and Icarus Verilog~\cite{Williams:2006}, as well as the commercial simulator ModelSim, using an agentic bug reporting tool that automatically classifies and files detected discrepancies.
\end{itemize}

\section{Background and Related Work}
\subsection{RTL Verification and Tool Reliability}
The reliability of EDA tools is a critical yet under-researched aspect of hardware design. Traditional verification focuses on the DUT, assuming the toolchain is correct. However, bugs in simulators can lead to "silent" failures where a design passes simulation but fails in silicon. Recent work on Verilog compiler cross-checking by Zhou et al. \cite{Zhou2025} and FPGA synthesis tool verification by Herklotz and Wickerson \cite{Herklotz2020} has highlighted the prevalence of bugs in even widely used tools. Differential testing, which compares the outputs of multiple tools on the same input, remains the most effective way to identify these discrepancies without a formal specification of the simulator's behavior. TransFuzz \cite{Solt2026} further demonstrated how translation-based fuzzing can reveal "confused deputy" attacks and other vulnerabilities in EDA software.

\subsection{BNF and Grammar-Based Fuzzing}
Backus-Naur Form (BNF) is a formal notation for describing context-free grammars, originally introduced by Naur \cite{Naur1960} for defining the ALGOL 60 programming language and later refined by Knuth \cite{Knuth1964}. BNF has become the de facto standard for specifying the syntax of programming and hardware description languages, including Verilog and SystemVerilog. A BNF grammar defines a set of production rules that specify how language constructs can be hierarchically combined.

Grammar-based fuzzing leverages these formal grammars to systematically generate syntactically valid test inputs by traversing production rules \cite{Sarkar2020}. Unlike random mutation-based fuzzing, grammar-based approaches guarantee that all generated inputs conform to the language specification, making them particularly suitable for testing parsers and compilers. Tools like Grammarinator \cite{Hodovan2018} and Gramatron \cite{Srivastava2021} have demonstrated the effectiveness of grammar-based approaches for discovering bugs in software parsers. Hardware fuzzing has evolved from simple random testing to sophisticated coverage-guided techniques. RFUZZ \cite{Laeufer2018} introduced coverage-directed fuzzing for RTL, while DifuzzRTL \cite{Hur2026} focused on discovering CPU bugs through differential testing. Recent tools like ChiGen \cite{Vieira2026} use bottom-up generation with probabilistic grammars and type inference to test EDA tools, outperforming earlier fuzzers like Verismith and VlogHammer. Despite these advances, generating high-quality RTL that is both syntactically valid and semantically interesting remains a challenge.

\subsection{LLMs for RTL Code Generation}
The application of LLMs to hardware design has seen rapid growth. Early works like VeriGen~\cite{Thakur2024} and the VerilogEval benchmark~\cite{Pinckney2025VerilogEval} demonstrate the feasibility of LLMs for generating functional Verilog code. However, generating syntactically and functionally correct RTL code remains challenging. To address this, recent research focuses on closed-loop feedback and verification-driven generation. For instance, RTLFixer~\cite{Tsai2024RTLFixer} employs RAG and ReAct prompting to automatically fix syntax errors using compiler feedback. Similarly, AutoVeriFix~\cite{Tan2026AutoVeriFix} uses a two-stage approach and simulation discrepancies to iteratively correct functional errors in generated Verilog. While these approaches significantly improve the syntactic and functional correctness of generated RTL, their primary objective is to produce spec-compliant code rather than stress-test the underlying toolchain.

\section{The \texttt{GI-RTL-Fuzz} Framework}
\label{sec:framework}

The primary objective of \texttt{GI-RTL-Fuzz} is to bridge the gap between stochastic grammar-based fuzzing and semantically-aware RTL generation. Traditional fuzzers like Verismith~\cite{Herklotz2020} generate RTL by randomly traversing production rules. While this ensures syntactic validity, it often fails to exercise deep semantic logic within simulators because: (1) the traversal is unguided, leading to high coverage of ``easy'' grammar rules and zero coverage of complex ones; and (2) it lacks the architectural intuition required to build meaningful hardware modules, resulting in RTL that is far removed from industrial applications.

\begin{figure}[t]
  \centering
  \includegraphics[width=0.95\linewidth]{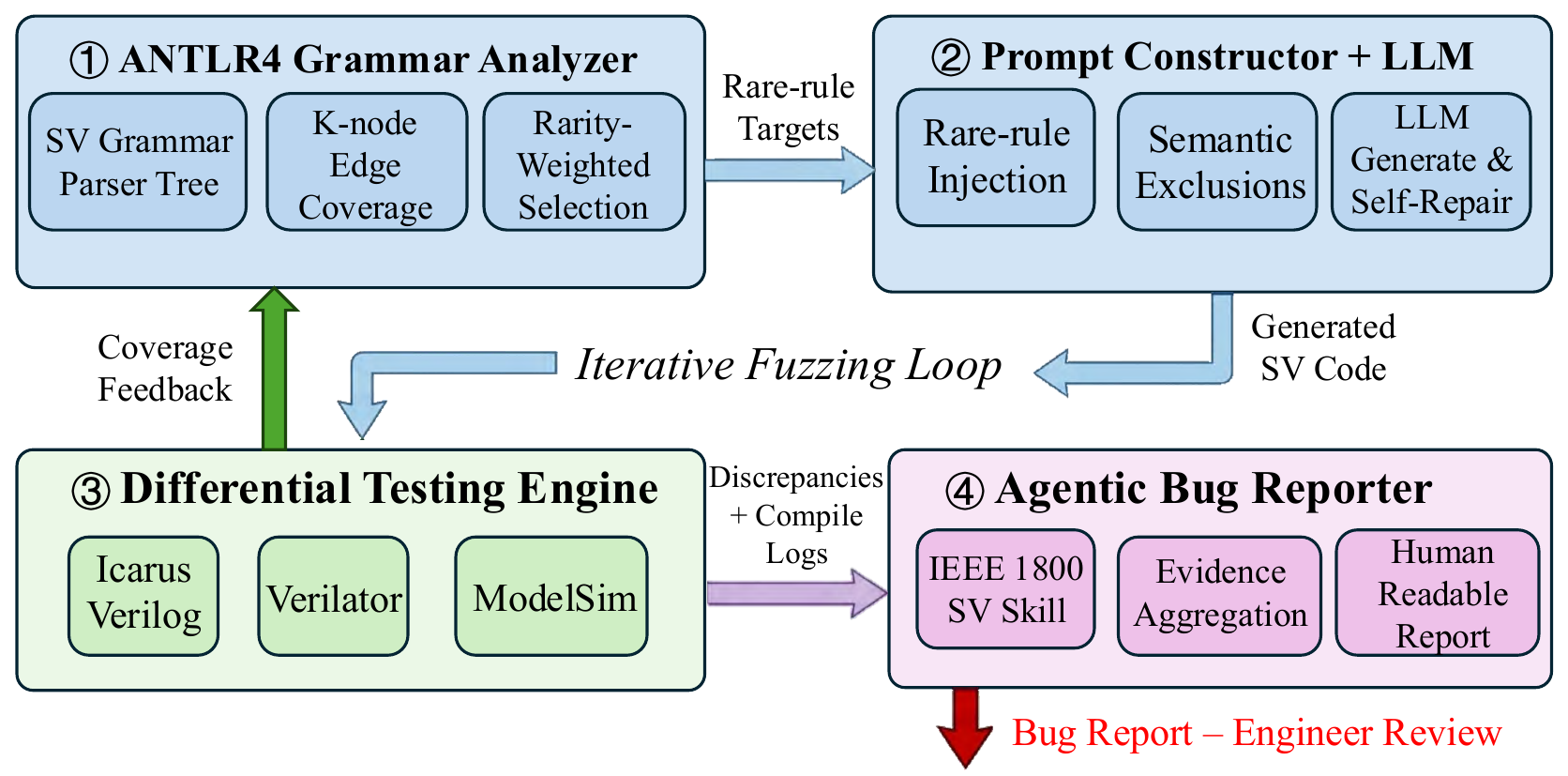}
  \caption{Overview of the \texttt{GI-RTL-Fuzz} framework.}
  \label{fig:architecture}
\end{figure}

\texttt{GI-RTL-Fuzz} addresses these limitations by using LLMs as semanticlly aware generators, guided by a \textit{Coverage-Aware Weighting} mechanism and a custom ANTLR4-based AST analyzer. The framework operates as a closed-loop system that iteratively increases grammar coverage while simultaneously discovering behavioral discrepancies among simulators. As illustrated in Figure~\ref{fig:architecture}, the system consists of four key components working together in an iterative loop:  
\begin{enumerate}
    \item \textbf{ANTLR4 Grammar Analyzer} parses the SystemVerilog grammar and measures $k$-node coverage to select underrepresented grammar targets.
    \item \textbf{Prompt Constructor and LLM Generation} module translates these targets into structured prompts with semantic constraints, invokes LLMs for code generation, and performs iterative self-repair based on linting feedback until a valid RTL module and testbench is produced.
    \item \textbf{Differential Testing Engine} then compiles and simulates the generated RTL across multiple simulators (e.g., Icarus Verilog, Verilator, and Modelsim), detecting both lint inconsistencies and simulation output mismatches at cycle-accurate granularity.
    \item \textbf{Agentic Bug Reporter} classifies and documents discrepancies by grounding findings in the SystemVerilog standard, producing structured bug reports for engineer review. Coverage metrics from the bug reporter are fed back to the Grammar Analyzer to guide subsequent iterations toward unexplored grammar regions.
\end{enumerate}

\subsection{ANTLR4 Grammar Analyzer}
\label{sec:grammar-analyzer}

The first component is responsible for parsing the SystemVerilog grammar, tracking coverage of grammar constructs, and selecting the next generation target. Our methodology employs a custom ANTLR4-based analyzer to perform a full recursive traversal of the SystemVerilog Abstract Syntax Tree (AST). Simple keyword counters or regular expression matchers often fail to handle the lexical complexity of SystemVerilog; therefore, our analyzer operates directly on the parse tree nodes.

We model coverage using \textit{$k$-node contexts}, each of which contains $k$ connected nodes in the AST. Tracking $k$-node contexts provides a richer representation of the RTL structural context and rule derivation history than simple node-based coverage. For example, considering a simplified ANTLR4 grammar for a conditional statement:

\begin{verbatim}
cond_stmt
  : 'if' '(' expr ')' stmt ('else' stmt)?
stmt
  : assignment | ';'
\end{verbatim}

A basic node-based coverage metric might only register the presence of an \texttt{if} statement. In contrast, by recording a $k$-node context, the analyzer captures the specific structural derivation. It can distinguish between an \texttt{if} branch leading to an assignment (the node sequence \texttt{cond\_stmt} $\to$ \texttt{stmt} $\to$ \texttt{assignment}) versus an \texttt{if} branch leading to an empty statement (the node sequence \texttt{cond\_stmt} $\to$ \texttt{stmt} $\to$ \texttt{';'}). This granularity is essential because simulator bugs frequently reside in the handling of specific branching logic or empty statements during elaboration. Similarly, for assignment contexts, $k$-node contexts allow the analyzer to differentiate between a procedural assignment within a block and a continuous assignment at the module level.

The analyzer implements a visitor pattern that recursively explores every node in the parse tree. For each node, it records the $k$-node context formed by the node and its ancestors. This structural information is aggregated into a global frequency counter $C(e)$ for each $k$-node context $e \in \mathcal{E}$. To prioritize rare constructs, we calculate a selection weight:
\[
  W(e) = \bigl(1 + C(e)\bigr)^{-1}
\]
. Contexts are then clustered by their parent non-terminal $P$, and a clustering score $S(P) = \sum_{e \in E_P} W(e)$ measures the collective exploration potential of each parent. Roulette-wheel sampling selects a high-value parent $P_{\text{target}}$, and the top $q$ most under-covered contexts under that parent are extracted as mandatory generation targets, where $q$ is the target quota per iteration.

\subsection{Prompt Constructor and LLM Generation}
\label{sec:prompt-llm}

\begin{figure}[t]
  \centering
  \includegraphics[width=0.9\linewidth]{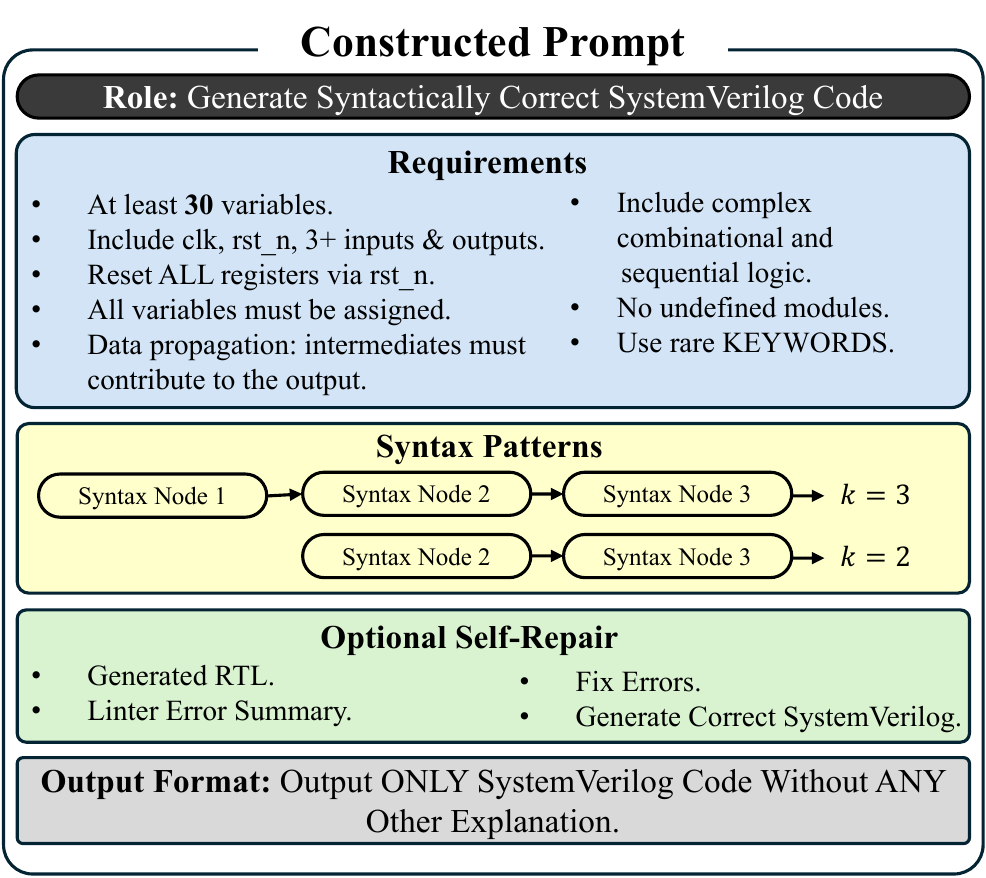}
  \caption{Composition of the grammar-informed LLM prompt.}
  \label{fig:prompt-workflow}
\end{figure}

The second component translates the rare grammar targets identified by the analyzer into structured prompts that guide the LLM to generate syntactically diverse and semantically meaningful SystemVerilog code. As illustrated in Figure~\ref{fig:prompt-workflow}, the prompt is composed of multiple sections that specify the generation task, grammar targets, synthesis constraints, and expected output format.

The prompt construction process consists of three stages: target transformation, structured prompt assembly, and iterative self-repair.

\textbf{Target Transformation.} Given the top-$q$ under-covered $k$-node contexts selected by the Grammar Analyzer, the Prompt Constructor transforms each node sequence (e.g., \texttt{A -> B -> C}) into a human-readable description by reversing the sequence and joining components with the keyword \texttt{inside} (e.g., ``implement C inside B inside A''). This transformation provides the LLM with clearer semantic intent about where the construct should appear in the module hierarchy. The mandatory block is then embedded into the prompt under the section header ``Advanced Syntax Challenges,'' which explicitly instructs the LLM to ``trigger corner-case simulator behaviors by nesting specific SystemVerilog constructs'' while ensuring the structures are ``semantically meaningful and logically connected to the module's primary data flow.''

\textbf{Structured Prompt Assembly.} As illustrated in Figure~\ref{fig:prompt-workflow}, the constructed prompt is organized into five sections. The Role section defines the generation task as producing a complete, syntactically correct SystemVerilog module. The Task Description outlines the primary objective and constraints. The Requirements section specifies synthesis requirements to ensure meaningful, non-degenerate RTL, including minimum variable counts, mandatory control signals (\texttt{clk} and \texttt{rst\_n}), diverse input naming conventions, required reset initialization, prohibition of floating signals, enforcement of data propagation where all intermediate variables contribute to outputs, and inclusion of rare SystemVerilog keywords (e.g., \texttt{IFNONE}, \texttt{FOREVER}, \texttt{PRIMITIVE}, \texttt{SPECIFY}) that frequently expose simulator edge cases. The Syntax Pattern section embeds the transformed $k$-node grammar targets as mandatory structural requirements for the LLM to exercise specific syntactic derivations. Finally, the Output Format section instructs the LLM to output only valid SystemVerilog code without any additional commentary or markdown formatting.

\textbf{Iterative Self-Repair Loop.} The generation pipeline operates as a closed-loop self-repair mechanism. After the LLM generates SystemVerilog code, the framework invokes a linter (e.g., Verilator in lint-only mode) to validate syntactic and semantic correctness. If errors are detected---such as \texttt{WIDTH} mismatch, \texttt{MULTIDRIVEN} variable conflicts, or syntax errors---the precise diagnostic message is extracted and appended to the current code as a repair prompt. The prompt instructs the LLM to perform a targeted fix while preserving the mandatory $k$-node grammar targets that were specified in the original prompt. This preserves the structural intent of the generation while correcting implementation errors. The process repeats until either the linter reports no errors or the maximum retry limit $R$ is reached. If the limit is exceeded, the module is discarded, and a new generation cycle begins. This closed-loop repair mechanism ensures a high yield of complex, compilable test cases that exercise the targeted grammar constructs.

\begin{algorithm}[t]
\small
\caption{Grammar-Informed RTL Generation}
\label{alg:generation}
\begin{algorithmic}
\REQUIRE Global counts $C$, $k$-node context size $k$, target quota $q$, max retries $R$, min variables $V$, total target $N$
\ENSURE Set of valid RTL modules $\mathcal{M}$
\STATE $\mathcal{M} \gets \emptyset$
\WHILE{$|\mathcal{M}| < N$}
    \STATE $W(e) \gets (1 + C(e))^{-1}$ \COMMENT{Calculate exploration weights}
    \STATE $S(P) \gets \sum_{e \in E_P} W(e)$ \COMMENT{Cluster weights by parent}
    \STATE Select $P_{\text{target}}$ via $\Pr(P) \propto S(P)$
    \STATE $\mathcal{T} \gets \text{Top-}q \text{ contexts in } E_{P_{\text{target}}} \text{ by } W$
    \STATE $\mathit{Prompt} \gets \text{ConstructPrompt}(P_{\text{target}}, \mathcal{T}, V)$
    \STATE $\mathit{Code} \gets \text{LLM.Generate}(\mathit{Prompt})$
    \STATE $r \gets 0$
    \WHILE{$\text{Lint}(\mathit{Code}) \neq \text{Pass} \land r < R$}
        \STATE $\mathit{Code} \gets \text{SelfRepair}(\mathit{Code})$
        \STATE $r \gets r + 1$
    \ENDWHILE
    \IF{$\text{Lint}(\mathit{Code}) \neq \text{Pass}$}
        \STATE \textbf{continue}
    \ENDIF
    \STATE $\mathit{Edges} \gets \text{ParseAST}(\mathit{Code})$
    \IF{$\mathcal{T} \subseteq \mathit{Edges}$}
        \STATE $\mathcal{M} \gets \mathcal{M} \cup \{\mathit{Code}\}$
        \STATE $C(e) \gets C(e) + 1$ for all $e \in \mathit{Edges}$
    \ENDIF
\ENDWHILE
\RETURN $\mathcal{M}$
\end{algorithmic}
\end{algorithm}

\subsection{Differential Testing Engine}
\label{sec:diff-testing}

The third component compiles and simulates each generated RTL module across multiple simulators and identifies behavioral discrepancies. The differential testing process consists of four stages: testbench generation, parallel simulation, VCD parsing, and trace comparison.

\textbf{Testbench Generation and Stimuli Construction.} For each generated RTL module, the framework automatically constructs a testbench by parsing the module's port interface. The parser handles both ANSI-style (input/output declared in module header) and non-ANSI-style (ports declared separately) module declarations, extracting signal names, directions (input/output), and bit-widths. The generated testbench includes: (1) clock generation using a \texttt{forever} loop with a fixed period (e.g., \texttt{\#5 clk = \textasciitilde clk}); (2) a reset sequence that drives \texttt{rst\_n} low for a configurable number of cycles; and (3) deterministic input stimuli driven on the falling edge to ensure signal stability at the rising edge. The stimuli are generated using a fixed random seed to ensure reproducibility across simulator runs.

\textbf{Parallel Simulation.} The same testbench and RTL module are compiled and simulated under multiple simulators, including Icarus Verilog, Verilator, and ModelSim. Each simulator produces a Value Change Dump (VCD) file containing the full waveform of all signals throughout the simulation. The framework executes simulations in parallel to minimize total wall-clock time.

\textbf{VCD Parsing and Trace Extraction.} The VCD comparison engine parses each simulator's VCD output using a dedicated parser that: (1) identifies all input and output signals of interest based on the parsed port list; (2) locates signal references within the hierarchical VCD scope (handling variations in scope naming such as \texttt{tb\_module}, \texttt{dut}, or flat hierarchies); and (3) extracts signal values at each rising clock edge. For multi-bit signals, the parser reconstructs the full bus value by combining individual bit signals in the correct bit order (high to low). The result is a trace consisting of $N$ cycles, where each cycle contains a snapshot of all input and output signal values at that clock edge.

\textbf{Trace Alignment and Comparison.} Before comparing traces from different simulators, the comparison engine applies a trace alignment algorithm that handles minor timing differences. The alignment shifts one trace relative to the other (in the range of $\pm 5$ cycles) and selects the shift that minimizes input mismatches, thereby accounting for differences in how simulators handle initial cycles or clock skew. After alignment, the comparison proceeds cycle-by-cycle and signal-by-signal. For each cycle, every output signal is compared across simulators. Unknown values (\texttt{x} or \texttt{z}) are handled specially: if both simulators report the same unknown value, the comparison is considered a match; if one reports a known value (\texttt{0} or \texttt{1}) and the other reports unknown, the result is marked as \texttt{UNKNOWN} rather than an error. Any divergence between known binary values (\texttt{0} vs. \texttt{1}) constitutes a \textit{simulation output mismatch}.

Two categories of discrepancy are detected. First, \textit{lint inconsistencies}: the compile-time diagnostic messages produced by each simulator are compared. A module that compiles cleanly under one simulator but triggers errors or warnings under another may indicate a non-conformant implementation of the language standard. Second, \textit{simulation output mismatches}: the VCD waveforms are compared at cycle-accurate granularity. The comparison engine normalizes signal representations into a unified binary format, aligns traces on the rising edge of the global clock, and flags any signal whose value diverges across simulators at any simulation cycle.

\subsection{Agentic Bug Reporter}
\label{sec:bug-reporter}

The fourth component automates the classification and documentation of detected discrepancies, transforming raw differential testing results into actionable bug reports suitable for simulator maintainers. We summarize the IEEE 1800-2017 SystemVerilog standard into a structured knowledge base, which serves as a \textit{skill} for the Agentic Bug Reporter to reference. This enables the agent to reason about whether a given behavioral divergence stems from an RTL design error or a genuine simulator deviation from the language specification.

The Agentic Bug Reporter receives three categories of inputs from the Differential Testing Engine: (1) the complete compile logs from each simulator under test; (2) the cycle-accurate VCD waveform differences pinpointing the exact cycle and signal at which simulators diverge; and (3) the generated RTL source code that triggered the discrepancy. An aggregator component scans for scenarios where simulators produce mismatched results and combines all relevant files and diagnostic information into a structured prompt that is fed to the agent. By invoking its IEEE 1800 skill, the agent retrieves pertinent clauses of the standard to assess compliance and identify root causes. It subsequently classifies each failure as either a lint inconsistency or a simulation output mismatch.

\begin{table}[t]
  \centering
  \caption{Example bug report from the Agentic Bug Reporter.}
  \label{tab:bug-report-example}
  \begin{tabular}{|p{0.95\linewidth}|}
    \hline
    \textbf{Bug:} SB12 --- \texttt{always\_comb} ignores textual statement execution order \\
    \hline
    \textbf{IEEE Clause:} IEEE Std 1800-2017, \S9.2.2.2 \\
    \textit{``Statements inside \texttt{always\_comb} shall execute in the order written.''} \\
    \hline
    \textbf{Affected Simulator:} Verilator \\
    \textbf{Conforming Simulators:} Icarus Verilog, ModelSim \\
    \hline
    \textbf{Example:} \texttt{result = a + accum; accum = b;} --- \texttt{result} should use old \texttt{accum} \\
    \hline
    \textbf{Severity:} Critical --- produces incorrect simulation results without warning. \\
    \hline
  \end{tabular}
\end{table}

These structured reports are intended for final review by human engineers, who confirm the findings and may submit them upstream to the simulator project maintainers. The feedback loop closes when coverage metrics from confirmed bugs are used by the Grammar Analyzer to prioritize generation targets that exercise the under-tested grammar regions.

\begin{table*}[htbp]
\caption{Bugs Discovered by \texttt{GI-RTL-Fuzz} in Simulators}
\centering
\vspace{-0.1in}
\renewcommand{\arraystretch}{1.2}
\resizebox{\textwidth}{!}{
\begin{tabular}{|c|c|c|l|c|c|c|}
\hline
\textbf{Bug} & \textbf{Simulator} & \textbf{Category} & \textbf{Description} & \textbf{Issue No.} & \textbf{Cross-checked} & \textbf{New?} \\
\hline

SB1  & Verilator & Parser & Binary use of unary reduction NAND (\texttt{\textasciitilde\&}) accepted & \#7353 & Verilator $\neq Icarus = ModelSim$ & \checkmark \\
\hline
SB2  & Verilator & Parser & Binary use of unary reduction NOR (\texttt{\textasciitilde|}) accepted & \#7353 & Verilator $\neq$ Icarus = ModelSim & \checkmark \\
\hline
SB3  & Verilator & Parser & Incorrectly accepts \texttt{!|} without parentheses & \#7417 & Verilator $\neq$ Icarus = ModelSim & \checkmark \\
\hline
SB4  & Verilator & Parser & Incorrectly accepts \texttt{!\&} without parentheses & \#7417 & Verilator $\neq$ Icarus = ModelSim & \checkmark \\
\hline
SB5  & Verilator & Parser & Incorrectly accepts \texttt{!\textasciicircum} without parentheses & \#7417 & Verilator $\neq$ Icarus = ModelSim & \checkmark \\
\hline
SB6  & Icarus    & Parser & Rejects valid labeled \texttt{begin-end} block & \#1321 & Icarus $\neq$ Verilator = ModelSim & \checkmark \\
\hline
SB7  & ModelSim  & Parser & Multiple continuous drivers on \texttt{logic} detected only at runtime & N/A & Icarus = Verilator $\neq$ ModelSim & \checkmark \\

\hline
SB8  & Verilator & Parser & Bare named block at module scope accepted outside generate & \#7349 & Verilator $\neq$ Icarus = ModelSim & \checkmark \\
\hline
SB9  & Icarus    & Semantic & \texttt{always\_ff} variable also written by \texttt{initial} & \#1317 & Icarus = Verilator $\neq$ ModelSim & \checkmark \\
\hline
SB10 & Verilator & Semantic & \texttt{always\_ff} variable also written by \texttt{initial} & \#7351 & Icarus = Verilator $\neq$ ModelSim & \checkmark \\
\hline
SB11 & Verilator & Semantic & Mixed declaration initialization and continuous assignment & \#7352 & Verilator $\neq$ Icarus = ModelSim & \checkmark \\

\hline
SB12 & Verilator & Simulation & \texttt{always\_comb} ignores textual execution order & \#7350 & Verilator $\neq$ Icarus = ModelSim & \checkmark \\
\hline
SB13 & Verilator & Simulation & Declaration initialization treated as time-varying & \#7348 & Verilator $\neq$ Icarus = ModelSim & \checkmark \\
\hline
SB14 & ModelSim  & Simulation & No warning on out-of-bounds array access & N/A & Icarus $\neq$ Verilator $\neq$ ModelSim & \checkmark \\
\hline
SB15 & Icarus    & Simulation & No warning on out-of-bounds array access & \#1322 & Icarus $\neq$ Verilator $\neq$ ModelSim & \checkmark \\
\hline
SB16 & Verilator & Simulation & Incorrect out-of-bounds array write behavior & \#7414 & Icarus $\neq$ Verilator $\neq$ ModelSim & $\times$ \\

\hline
\end{tabular}%
}
\label{tab:simulator_bugs}
\vspace{-0.3cm}
\end{table*}

\section{Experimental Evaluation}
\label{sec:experiments}

\subsection{Experimental Setup}
We evaluated GI-RTL-Fuzz by generating 1000 SystemVerilog test cases using GPT-4-0125, each requiring a minimum of 30 variables to ensure structural complexity. The designs were tested against Icarus Verilog (v13.0), Verilator (v5.046), and ModelSim (v2020.4).

All experiments were conducted on a high-performance computing cluster equipped with Intel Xeon Gold 6348 CPUs and 512GB RAM. This computational setup enabled parallel execution of simulator testing across multiple cores, significantly reducing the total wall-clock time required for differential testing of 1000 generated RTL modules against three distinct simulators. Each simulator was invoked with strict warning levels enabled to ensure that any diagnostic messages were properly reported. The agentic bug reporting tool, implemented using KiloCode, automatically records and files detected discrepancies by leveraging an LLM with reference to the IEEE 1800-2017 standard.

\subsection{Bugs Found in Verilog Simulators}
\label{sec:simulator_bugs}

To evaluate the effectiveness of \texttt{GI-RTL-Fuzz} in uncovering defects in real-world Verilog and SystemVerilog simulators, we applied it to open-source simulators (Icarus Verilog and Verilator) and one commercial simulator (ModelSim). Table~\ref{tab:simulator_bugs} summarizes the bugs discovered. In total, \texttt{GI-RTL-Fuzz} uncovered 16 bugs, \textbf{15} of which were previously unknown bugs, spanning three categories: parser, semantic analysis violations, and incorrect simulation behavior. All reported bugs violate the SystemVerilog standard, and each was cross-checked against at least two other simulators.

\subsubsection{Parser and Front-End Bugs}

SB1 and SB2 expose closely related parser defects in Verilator involving misuse of unary reduction operators. IEEE~1800 defines reduction NAND (\texttt{\textasciitilde\&}) and reduction NOR (\texttt{\textasciitilde|}) as unary operators. However, Verilator incorrectly accepts them in binary form, e.g., \texttt{a \textasciitilde\& b} or \texttt{a \textasciitilde| b}. The correct forms must use unary reduction (e.g., \texttt{\textasciitilde\&a}) or explicit binary logic (e.g., \texttt{\textasciitilde(a \& b)}). Icarus Verilog and ModelSim correctly reject these constructs.

SB3, SB4, and SB5 reveal additional parsing inconsistencies in Verilator related to operator precedence. Expressions such as \texttt{!|a}, \texttt{!\&a}, and \texttt{!\textasciicircum a} are accepted without parentheses, even though IEEE~1800 requires explicit disambiguation (e.g., \texttt{!(|a)}). Icarus Verilog and ModelSim reject these malformed expressions, indicating that Verilator’s precedence handling is overly permissive.

SB6 is a parser compliance issue in Icarus Verilog. It rejects a valid statement-labeled block written as \texttt{label: begin ... end}, while accepting \texttt{begin: label ... end}. IEEE~1800 permits both forms, and both Verilator and ModelSim accept them, indicating incomplete grammar support in Icarus.

SB7 is a ModelSim front-end checking issue. The simulator accepts multiple continuous assignments to a \texttt{logic} variable during compilation, but reports the violation only at runtime. IEEE~1800 (\S6.5, \S10.5) requires such errors to be detected earlier. Both Verilator and Icarus correctly reject the construct during compilation.

SB8 reveals an elaboration defect in Verilator involving named blocks at module scope. Verilator accepts a bare named block outside of a valid \texttt{generate} construct, violating IEEE~1800 rules governing module items and generate regions (\S23, \S27). Both Icarus Verilog and ModelSim correctly reject this construct.

\begin{table*}[t]
  \caption{Detailed Breakdown of Simulation Outcomes for GPT-4 across $k$-node Configurations (1000 Units per Configuration).}
  \label{tab:knode_coverage}
  \centering
  \small
  \begin{tabular}{lccccc}
    \toprule
    \textbf{Generator} & \textbf{$k$-node} & \textbf{Success} & \textbf{$k$=3 Node Coverage} & \textbf{Compilation} & \textbf{Behavioral} \\
    & \textbf{Config} & \textbf{Units} & \textbf{(\%)} & \textbf{Mismatch (\#)} & \textbf{Mismatch (\#)} \\
    \midrule
    \textbf{GPT-4} & $k$=1 & 815 & 43.05\% & 77 & 30 \\
    \textbf{} & $k$=2 & 710 & 48.23\% & 75 & 36 \\
    \textbf{} & $k$=3 & 806 & 51.92\% & 92 & 41 \\
    \midrule
    \textbf{GPT-4 Baseline} & N/A & 862 & 35.12\% & 0 & 23 \\
    \midrule
    \textbf{Verismith} & N/A & 1000 & 11.08\% & 0 & 0 \\
    \bottomrule
  \end{tabular}

\end{table*}

\subsubsection{Semantic Analysis Bugs}

SB9 and SB10 expose the same semantic defect in Icarus Verilog and Verilator, respectively. In both simulators, a variable assigned inside an \texttt{always\_ff} block is also allowed to be written by an \texttt{initial} block. This violates IEEE~1800 \S9.2.2.4, which requires that variables driven by \texttt{always\_ff} have a single writer. ModelSim correctly enforces this restriction.

SB11 is a Verilator-specific semantic bug involving conflicting driver models. A \texttt{logic} variable is both declaration-initialized and driven by a continuous assignment. Since declaration initialization is procedural (\S6.5) and \texttt{assign} introduces a continuous driver (\S10.5), the combination is illegal. Icarus Verilog and ModelSim reject the construct, while Verilator accepts it.

\subsubsection{Simulation Bugs}

SB12 and SB13 are among the most severe classes of defects identified by \texttt{GI-RTL-Fuzz}, because they produce incorrect simulation results without diagnostic errors.

SB12 demonstrates that Verilator does not preserve textual execution order within an \texttt{always\_comb} block. IEEE~1800 \S9.2.2.2 requires sequential execution semantics. In the example:
\vspace{-0.04in}
\begin{verbatim}
    result = a + accum;
    accum  = b;
\end{verbatim}
\texttt{result} must observe the previous value of \texttt{accum}. However, Verilator collapses the block into a dependency graph and computes \texttt{result = a + b}, violating the standard. Icarus Verilog and ModelSim behave correctly.

SB13 exposes incorrect handling of declaration-time initialization. A declaration such as \texttt{logic [7:0] snapshot = data\_in;} should be evaluated once at time zero (\S6.5, \S10.5). Verilator instead treats it as time-varying, effectively behaving like a continuous assignment.

SB14 and SB15 reveal missing diagnostics for out-of-bounds array accesses in ModelSim and Icarus, respectively. In both cases, illegal indexing does not trigger warnings.

SB16 exposes a more severe issue in Verilator: incorrect handling of out-of-bounds array writes. Instead of merely lacking diagnostics, the simulator produces incorrect state updates, leading to silent corruption of the simulation state.

Overall, these results demonstrate that \texttt{GI-RTL-Fuzz} is capable of uncovering a broad spectrum of defects in industrial-strength HDL simulators, ranging from front-end parsing issues to deeper semantic checking failures and silent simulation miscomputations. Among these, the simulation bugs are especially concerning because they do not merely reject or misclassify invalid code; instead, they produce plausible but incorrect execution results without warning, potentially leading users to trust flawed simulation outcomes during RTL verification. \vspace{-0.3cm}

\subsection{Grammar Coverage Analysis}
\label{sec:grammar_coverage}

By explicitly targeting rare BNF transitions, \texttt{GI-RTL-Fuzz} successfully synthesized complex constructs such as nested \texttt{generate} blocks with conditional \texttt{specify} paths, which are traditionally difficult for random fuzzers to produce while maintaining syntactic validity.

We measure grammar coverage as $k$-node coverage, as described in Section~\ref{sec:grammar-analyzer}. The $k$-node context size determines the granularity of grammar coverage: $k$=1 tracks individual nodes (331 possible combinations), $k$=2 captures parent-child contexts (641 possible combinations), and $k$=3 captures grandparent-parent-child contexts (1273 possible combinations). Our ablation study demonstrates that $k$=3 provides the best trade-off between coverage granularity and bug discovery effectiveness.

To further benchmark the effectiveness of grammar-informed generation, we compared \texttt{GI-RTL-Fuzz} with Verismith \cite{Herklotz2020}, a state-of-the-art grammar-based RTL generator. As shown in Table~\ref{tab:knode_coverage}, Verismith produced 1000 compilable units but achieved only 11.08\% $k$=3 node coverage and discovered zero simulator discrepancies. In contrast, \texttt{GI-RTL-Fuzz} guided GPT-4 to 51.92\% $k$=3 node coverage and discovered 41 behavioral mismatches. This stark contrast demonstrates that our method successfully directs the LLM to systematically explore the vast, under-tested regions of the SystemVerilog grammar that are unreachable by conventional grammar-based generation.

The key differentiator lies in our coverage-aware weighting mechanism. Unlike Verismith, which relies on random traversal of production rules without semantic guidance, \texttt{GI-RTL-Fuzz} prioritizes underrepresented grammar edges based on their historical coverage frequency. This ensures that the LLM is prompted to generate constructs from grammar regions that are rarely exercised in typical hardware designs, thereby increasing the likelihood of triggering corner-case simulator behaviors.

The baseline GPT-4 (labeled ``N/A'' in subsequent tables) refers to GPT-4 generation without grammar-informed guidance. Without explicit grammar targets, LLMs tend to converge on common patterns after generating a certain amount of code, failing to produce sufficiently diverse RTL designs. As shown in Table~\ref{tab:knode_coverage}, this baseline achieved 35.12\% $k$=3 node coverage with 23 behavioral mismatches, compared to 51.92\% $k$=3 node coverage with 41 behavioral mismatches for GPT-4 guided by $k$=3 grammar targets. The 78\% improvement in behavioral mismatch detection (23 vs 41) validates the effectiveness of our coverage-aware weighting mechanism in directing the LLM toward grammar regions that trigger simulator bugs.

\subsection{Impact of $k$-node Context Size on Bug Discovery}
To understand how different grammar context sizes affect bug discovery, we conducted an ablation study across $k$=1, $k$=2, and $k$=3 configurations. As described in Section~\ref{sec:grammar-analyzer}, $k$-node context size determines the granularity of grammar coverage. Larger $k$-node contexts capture more complex grammatical relationships, enabling the discovery of bugs that reside in deeply nested construct interactions.

Table~\ref{tab:bug_detection_matrix} presents the bug detection results across different configurations. The baseline without $k$-node guidance detected 7 bugs, while GPT-4 with $k$=3 guidance detected 15 bugs. Notably, bugs SB14, SB15, and SB16 were exclusively discovered with $k$=3 guidance. These bugs involve out-of-bounds array access handling, which requires complex interaction between module instantiation, procedural blocks, and signal scheduling---precisely the type of deeply nested grammar constructs that $k$=3 tracking can expose.

We selected $k$=3 as the maximum context size based on the empirical observation that larger values would require tracking over 2500 distinct grammatical relationships, making coverage selection less focused and the exploration mechanism less efficient. The bug detection matrix confirms this design choice: $k$=3 achieves the highest bug detection rate while maintaining tractable coverage tracking.

\begin{table}[t]
  \caption{Bug Detection Matrix for GPT-4 across $k$-node Configurations. A checkmark ($\checkmark$) indicates that the corresponding bug was discovered under that configuration.  \vspace{-0.3cm}}
  \label{tab:bug_detection_matrix}
  \centering
  \small
  \begin{tabular}{lcccc}
    \toprule
    \textbf{Bug ID} & \textbf{Baseline} & \textbf{$k$=1} & \textbf{$k$=2} & \textbf{$k$=3} \\
    \midrule
    SB1  & \checkmark &        &        & \checkmark \\
    SB2  &        & \checkmark &        & \checkmark \\
    SB3  & \checkmark &        & \checkmark & \checkmark \\
    SB4  & \checkmark &        & \checkmark & \checkmark \\
    SB5  & \checkmark &        & \checkmark & \checkmark \\
    SB6  &        & \checkmark &        & \checkmark \\
    SB7  & \checkmark & \checkmark &        & \checkmark \\
    SB8  &        & \checkmark & \checkmark &        \\
    SB9  & \checkmark & \checkmark &        & \checkmark \\
    SB10 & \checkmark       & \checkmark &        & \checkmark \\
    SB11 &        & \checkmark & \checkmark & \checkmark \\
    SB12 &        & \checkmark &        & \checkmark \\
    SB13 &        & \checkmark &        & \checkmark \\
    SB14 &        &        &        & \checkmark \\
    SB15 &        &        &        & \checkmark \\
    SB16 &        &        &        & \checkmark \\
    \midrule
    \textbf{Total} & \textbf{7} & \textbf{9} & \textbf{5} & \textbf{15} \\
    \bottomrule
    \vspace{-1cm}
  \end{tabular}
\end{table}

The data reveals that $k$=3 guidance produces the broadest bug-discovery profile, identifying 15 of the 16 unique bugs represented in the matrix and exceeding all other configurations. Notably, Icarus Verilog accounts for the vast majority of single-simulator compilation failures, often due to its stricter (or sometimes incomplete) implementation of SystemVerilog's more advanced features.

\subsection{Discussion}
\label{sec:discussion}

The SystemVerilog standard (IEEE 1800) contains numerous behaviors that can lead to non-deterministic simulation results across different simulators. According to IEEE 1800 \S4.8, race conditions arise when the execution of expression evaluation and net update events may be intermingled, resulting in simulator-dependent behavior. The standard explicitly states that such races are permissible and that different simulators may correctly produce different results for the same code. Additionally, IEEE 1800 \S6.5 specifies that when a variable is written by multiple procedural statements, ``the last write determines the value,'' yet the actual ordering within a delta cycle is simulator-defined. While these behaviors may faithfully reflect actual hardware semantics, they pose significant challenges for verification engineers who rely on consistent simulator behavior. Critically, many of these constructs produce no warnings, making it difficult to identify potential issues during simulation.

We provide two case studies to illustrate common sources of behavioral mismatches. As shown in Table~\ref{tab:knode_coverage}, these mismatches are not necessarily indicative of simulator bugs---they often arise from behaviors permitted by the IEEE 1800 standard, where different simulation semantics are both considered correct.

\subsubsection*{Case Study 1: Multiple Procedural Assignments}
Consider the following code pattern where two procedural blocks assign to the same variable:

\begin{verbatim}
    always @(posedge clk) result <= data_a; // P1
    always @(posedge clk) result <= data_b; // P2
\end{verbatim}

Both assignments target the same variable \texttt{result} within the same clock edge. According to IEEE 1800 \S6.5, variables can be written by multiple procedural statements, and ``the last write determines the value.'' However, since both assignments occur simultaneously in a delta cycle, the final value depends entirely on the simulator's scheduling algorithm. Our testing reveals that none of the three simulators (Verilator, Icarus, ModelSim) issue any warning for this construct, despite its inherently non-deterministic behavior. This silence can mislead engineers into believing the simulation is deterministic when it is not.

\subsubsection*{Case Study 2: Scheduling Race Through a Continuous Assignment}
Consider the following code where a signal is computed via continuous assignment and used within the same clock edge:

\begin{verbatim}
    assign overflow = (data == 8'hFF);
    always @(posedge clk)
        data = data + 1;
    always @(posedge clk)
        if (overflow)
            flag = 1'b1;
\end{verbatim}
Here, \texttt{overflow} is computed from \texttt{data} through a continuous assignment. However, \texttt{data} is also updated on the same clock edge in a different \texttt{always} block. Whether \texttt{overflow} reflects the old or new value of \texttt{data} depends on the simulator's evaluation order, leading to non-deterministic behavior. 

One possible improvement would be for simulators to drive signals to \texttt{x} when such undefined behaviors create ambiguous execution order, helping engineers detect potential race conditions early in the verification process. Additionally, it may be beneficial for the  standard to mandate more explicit warnings for race conditions and undefined behaviors, requiring simulators to provide better error reporting to reduce behavioral mismatches during verification.
\vspace{0.4cm}

\vspace{-0.7cm}
\section{Conclusion}
In this paper, we introduced \texttt{GI-RTL-Fuzz}, a novel grammar informed RTL generation framework designed for differential fuzzing of RTL simulators. By combining the generative power of Large Language Models with BNF grammar coverage guidance, our framework successfully steered the generation process toward complex and rarely-tested language features. Our differential testing pipeline uncovered 15 previously unknown bugs across Icarus Verilog, Verilator, and ModelSim, spanning parser, semantic, and simulation categories.

The results highlight a critical reliability gap in the hardware toolchain: simulators often fail to warn users about violations of the IEEE 1800 standard, leading to silent functional mismatches between simulation and actual silicon. Additionally, we discussed how undefined and simulator-dependent behaviors permitted by the IEEE 1800 standard can result in non-deterministic simulation results across different tools, further complicating verification efforts. By systematically exposing these gaps, \texttt{GI-RTL-Fuzz} demonstrates the importance of grammar-informed fuzzing for improving the robustness of HDL simulation toolchains.

\bibliographystyle{ACM-Reference-Format}
\bibliography{references}

\end{document}